\documentclass[12pt]{article}
\usepackage{cite}
\usepackage{amssymb}
\usepackage{pifont}\usepackage{amsmath}
\usepackage{subfigure}
\usepackage{graphics}
\usepackage{pstricks}
\usepackage{pst-plot}
\usepackage{pst-slpe}
\def\eq#1{(\ref{#1})}

\def\pa{\partial}
\def\rt{\longrightarrow}

\def\viz{{\slshape viz.~}}

\newcommand\dual[1]{{#1}}

\newcommand\ternary[1]{{\langle\,#1\,\rangle}}

\newcommand\G{\Gamma}
\newcommand\g{\gamma}
\begin{document}
\title{BLG theory with generalized Jordan triple systems}
\author{
Sudipto Paul Chowdhury \thanks{tpspc@iacs.res.in} \\
\small Department of Theoretical Physics \&
\small Centre for Theoretical Sciences\\
\small Indian Association for
\small the Cultivation of Science\\
\small Calcutta 700 032, India\\ 
Subir Mukhopadhyay \thanks{subir@iopb.res.in} \\
\small NISER, Bhubaneshwar 751 005, India\\
Koushik Ray \thanks{koushik@iacs.res.in}\\
\small Department of Theoretical Physics \&
\small Centre for Theoretical Sciences\\
\small Indian Association for
\small the Cultivation of Science\\
\small Calcutta 700 032, India.
}
\date{}
\maketitle
\begin{abstract}
\thispagestyle{empty}
\noindent We use a generalized Jordan algebra of the second kind to
study the recently proposed BLG theory of multiple M2-branes.
We find the restriction imposed on the ternary product from
its consistency with the BLG theory.
\end{abstract}
\clearpage
\noindent 
The various string theories are believed to unify into the
eleven-dimensional M-theory,
based on strong circumstantial evidences. M-theory contains two
types of configurations in its spectrum, namely membranes and pentabranes, 
M2- and M5-branes, respectively, for short. These are envisaged as
strong-coupling manifestations of their ``stringy" counterparts, namely,
D2- and D5-branes, respectively. However, while the world-volume theories of
D-branes have been studied extensively from various points of
view, leading occasionally to previously unknown field theories, 
a world-volume description of M-branes have been lacking. 
Recently, the world-volume theory of coalescing M2-branes have
received a lot of attention. The approaches to obtaining a theory of
M2-branes may be broadly classified in two categories. The first is 
based solely on supersymmetry considerations and a modification of the
algebraic structure of the underlying gauge theory, 
thereby fixing all interactions
\cite{bagger1,bagger2,bagger3,bagger4,gustaf}. The other is
modelled after quiver gauge theories cashing in on  the AdS-CFT
correspondence \cite{abjm}. It is not unlikely that these two 
approaches are related
too. In this article we shall pursue the first approach, working with
a BLG theory, named after its inventors \cite{bagger1,gustaf}. 

Three attributes are taken to be the hallmark of the 
world-volume theory of multiple M2-branes. 
First, it is to have eight bosonic scalar fields which may
be interpreted as the eight directions transverse to the M2-branes
in the eleven-dimensional M-theory. secondly, one expects the theory 
to have sixteen supersymmetries, as in the theory of D2-branes. 
Thirdly, it is to emerge at the 
conformally invariant infra-red fixed point of the theory of coalescing
D2-branes. The BLG proposal satisfies these criteria.
The field content of the BLG theory are eight bosonic scalars,
fermions and gauge fields valued in a ternary algebra. Let us
remark in passing that the use of ternary or $n$-ary algebras for modelling
configurations in a strongly coupled situation is not unprecedented.
Such algebras have been
used in the context of hadronic Physics earlier \cite{nambu,
santilli}. A ternary algebra was used in such instances also 
to model three-body
interactions. While M2-branes are configurations obtained from
D2-branes in a strongly coupled regime, the algebra in this case 
is not used to describe three-body interactions. On the contrary, 
it is deemed to model BPS
configuration of multiple M2-branes. The use of a ternary algebra in
this context as
opposed to an $n$-ary one for $n>3$ is motivated by supergravity
considerations \cite{basu}. 

Various aspects of the proposed theory based on a ternary algebra has
been worked out
\cite{mukhi1,mukhi2,mukhi3,raam,figu1,figu2,pass,cherkis,
jabbari,singh,krishnan, palmkvist,Bonelli}. 
One of the hurdles in the construction of
the theory seems to be a choice of the ternary algebra itself. If the
algebra is characterized by completely anti-symmetric structure
constants satisfying a certain identity, called the fundamental identity,
which generalizes the Jacobi identity of Lie algebras, and is
required to have a Euclidean metric made out of bilinear invariants
of the algebra, then the only choice for structure constants
one is left with is
proportional to the rank four antisymmetric tensor. One can therefore 
only recover the
theory of \emph{two} D2-branes from it \cite{mukhi1,mukhi2}. 
Interesting variations of 
this theory by relaxing one or more of these requirements have been
considered, but a consensus on \emph{the}
theory of multiple M2-branes is being awaited. 
In this article we propose to use 
a generalized Jordan triple system of the second kind 
(GJTS-II) \cite{kamiya,kammon}, which is a ternary algebra which does
not use the fundamental identity. 

A GJTS-II is an algebra 
$\mathcal{A}=\{V,\ternary{\null}\}$
consisting of a finite-dimensional vector space $V$, over a field of
characteristic zero, chosen to be the field of real numbers here, endowed with 
a ternary product 
\begin{equation} 
\begin{split}
V\otimes V &\otimes V\rt V, \\
(A,B,C)&\mapsto\ternary{ABC},
\end{split}
\end{equation} 
obeying the identities
\begin{equation}  
\label{JK1}
\ternary{AB\ternary{CDE}}
-\ternary{\ternary{ABC}DE}
+\ternary{C\ternary{BAD}E}
-\ternary{CD\ternary{ABE}} =0,
\end{equation} 
\begin{equation} 
\label{JK2}
\begin{split}
\ternary{AE\ternary{CBD}}
&-\ternary{\ternary{CBD}EA}
+\ternary{AB\ternary{CED}}
+\ternary{C\ternary{BAE}D}\\
&\!\!\!\!\!\!-\ternary{AE\ternary{DBC}}
+\ternary{\ternary{DBC}EA}
-\ternary{AB\ternary{DEC}}
-\ternary{D\ternary{BAE}C}
 = 0,
\end{split}
\end{equation} 
where $A$, $B$, $C$, $D$, $E$ are elements of  $V$.
The identities \eq{JK1} and \eq{JK2} are
referred to, respectively, as Jacobson's and Kantor's identities. 
These identities guarantee the emergence of graded Lie algebras from a
GJTS-II. Indeed, 
\emph{all} compact and non-compact semi-simple
Lie algebras can be recovered
from different classes of GJTS-II 
upon choosing different vector spaces $V$ and appropirate ternary products
\cite{bars}.  The linear endomorphism 
of $\mathcal{A}$,
\begin{equation} 
L_{AB}(C) := \ternary{ABC},
\end{equation} 
satisfying 
\begin{equation}
\label{lder}
[L_{AB},L_{CD}]=L_{L_{AB}(C),D}-L_{C,L_{BA}(D)},
\end{equation} 
by the Jacobson's identity \eq{JK1},
defines
a derivation on the algebra $\mathcal{A}$ as 
\begin{equation} 
\label{derdef}
\mathcal{D}_{AB} := L_{AB}-L_{BA},
\end{equation} 
such that 
\begin{equation}
[\mathcal{D}_{AB},L_{CD}]=L_{\mathcal{D}_{AB}(C),D}+L_{C,\mathcal{D}_{AB}(D)},
\end{equation} 
as can be verified using \eq{lder}.
Let us remark that, had we required the endomorphism $L$ to be a
derivation with respect to the ternary product, 
then the corresponding Leibniz rule 
\begin{equation}
L_{AB}\ternary{CDE} =  \ternary{L_{AB}(C)DE} +
\ternary{CL_{AB}(D)E}+\ternary{CDL_{AB}(E)}
\end{equation} 
would have become the so-called ``fundamental identity",
\begin{equation} 
\ternary{AB\ternary{CDE}} =  \ternary{\ternary{ABC}DE} +
\ternary{C\ternary{ABD}E}+\ternary{CD\ternary{ABE}}.
\end{equation} 
This feature is different from the ``usual" gauge theories based on
Lie algebras, as well as the traditional version of the BLG theory,
where the gauge and matter fields belong to the same Lie triple
system. We do not require the ternary product to satisfy the
fundamental identity, unlike the traditional BLG theory. 
Indeed, combining the algebra $\mathcal{A}$ and the
derivations $\mathcal{D}$ together in a single set, 
one obtains a Lie triple system from the
GJTS-II, and thence a graded
Lie algebra, usually referred to as a Kantor algebra, 
with the identities \eq{JK1} and \eq{JK2}
enforcing the Jacobi identity of
the generators of the Lie algebra \cite{bars,kammon}.

A bilinear form, attributed to Yamaguti \cite{yamaguti}, on
a GJTS-II is defined as \cite{kamiya,kammon}
\begin{equation}
\label{tf}
\gamma(A,B) := \frac{1}{2}\mathrm{Tr}\left(
2R_{AB}+2R_{BA}-L_{AB}-L_{BA}\right), 
\end{equation} 
where the linear transformations $L$ and $R$ are defined as 
$L_{AB}(C)=\ternary{ABC}$ and $R_{AB}(C)=\ternary{CAB}$, and the
trace is taken over the linear endomorphisms. 
The bilinear form $\gamma$, known as the \emph{trace form}, 
coincides with the Killing form of the corresponding Kantor
algebra, up to normalization, 
provided $\mathcal{A}$ satisfies some extra condition. 

For future use
let us introduce the fully antisymmetrized combination of the ternary
brackets
\begin{equation}
\label{omegadef}
\Omega_{AB}(C) =
\ternary{ABC}
+\ternary{BCA}
+\ternary{CAB}
-\ternary{BAC}
-\ternary{ACB}
-\ternary{CBA}
\end{equation} 
$A$, $B$, $C$ being vectors in $V$.
This satisfies $\Omega_{AB}(C) = \Omega_{BC}(A)=\Omega_{CA}(B)$ and
$\Omega_{AB}(C) = -\Omega_{BA}(C)$.
We shall, whenever convenient, use a set of basis vectors of $V$ and
denote them by
$\tau$, so that $V=\{\tau_a\}_{\mathrm{span}}$, with the subscript $a$
belonging to an appropriate index set on $\mathcal{A}$.
We shall often abuse notation by labelling operators with the indices
$a$, $b$ and suppressing  $\tau$. For example, we shall use the symbols
$L_{ab}$ and  
$\mathcal{D}_{ab}$  for 
$L_{\tau_a\tau_b}$ and
$\mathcal{D}_{\tau_a\tau_b}$, respectively.

Let us now consider the BLG theory in terms of a GJTS-II,
beginning with the field content of the three-dimensional world-volume 
theory of M2-branes. The world-volume fills the directions $0,1,2$, which 
will be indicated by Greek letters.
The transverse directions furnish eight scalars to the world-volume 
gauge theory,
$X^I$, $I=3,\cdots, 10$. Correspondingly, there are eight $Spin(1,2)$
fermions in the world-volume collected together in a spinor field,
denoted $\psi$. Thus, $\psi$ is an eleven-dimensional Majorana spinor
with sixteen independent real components
satisfying $\G_{012}\psi = -\psi$. In addition, the closure of the
supersymmetry algebra calls for the incorporation of a vector field,
which is required to be non-dynamical \cite{schwarz} and hence chosen
to appear in the action as a Chern-Simons term only. 
We assume that the bosons $X^I$ as well as the fermion $\psi$ are
valued in the vector space $V$. Thus, with the assumption that
 $V$ to be the linear span
of a finite number of vectors $\tau$, as mentioned above, 
we write the fields as 
\begin{equation}
\begin{split}
X^I = \sum_{a} x^I_a\tau_a\\
\psi = \sum_{a} \psi_a\tau_a.
\end{split} 
\end{equation} 
The vector fields are taken to be valued in $Der(\mathcal{A})$ as
\begin{equation} 
\label{Amu}
A_{\mu} = \sum_{a,b}A_{\mu}^{ab}\mathcal{D}_{ab}. 
\end{equation} 
It follows from  \eq{derdef} that $A_{\mu}^{ab}$ is antisymmetric in
the indices $a$, $b$. We shall find that 
$A_{\mu}$ plays the role of gauge fields in the
sequel. 
The BLG action  is  written in terms of the bilinear trace form
\eq{tf} as
\begin{equation}
\label{terac}
\begin{split}
\mathcal{L} = \mathcal{L}_\mathrm{CS}
-\frac{1}{2} \gamma({D_{\mu}X^I}, {D^{\mu} X^I}) 
+\frac{i}{2}\gamma({\bar{\psi}}, {\g^{\mu}D_{\mu}\psi}) 
&- \frac{i}{4}\G^{IJ}\gamma({\bar{\psi}},\Omega_{X^IX^J}(\psi))
\\ &~~~- \frac{1}{12}\gamma(\Omega_{X^J X^K}(X^I),(\Omega_{X^J X^K}(X^I)),
\end{split}
\end{equation} 
where $\mu=0,1,2$ is an index on the 
three-dimensional world-volume, $\g^{\mu}$ are the
three-dimensional Gamma matrices, $\G^I$ are the eight-dimensional
Gamma matrices.
We defined the derivation $D_{\mu}$ on the fields as 
\begin{equation}
\label{deriv}
D_{\mu} \Phi = \pa_{\mu} \Phi - A_{\mu}(\Phi),
\end{equation} 
with the action of $A_{\mu}$ on a field $\Phi$ defined in accordance with 
\eq{Amu} as 
\begin{equation} 
\label{amu}
A_{\mu}(\Phi) = \sum_{a,b,c}A_{\mu}^{ab}\Phi^c\ternary{\tau_a\tau_b\tau_c}.
\end{equation} 
We shall find that $D_{\mu}$ acts as a covariant derivative for a
certain gauge transformation of the fields and we shall 
assign a transformation of $A_{\mu}^{ab}$ under the gauge
transformation, as we come across it in the sequel. For the time being
it suffices to treat it as a vector field in the theory.
The Chern-Simons term 
is defined as in the traditional BLG
theory \cite{bagger1}
\begin{equation}
\mathcal{L}_{\mathrm{CS}} = 
\frac{1}{2} \mathrm{Tr} \; {A} \wedge F,
\end{equation} 
where we defined $F_{\mu\nu}=[D_{\mu},D_{\nu}]$ and the trace is as in 
\eq{tf}.
The action $\mathcal{L}$ in \eq{terac} is invariant under the 
supersymmetry transformations \cite{bagger3},
\begin{gather} 
\label{susy}
\delta X^I = i\bar{\theta}\G^I\psi\\
\label{susypsi}
\delta \psi = D_{\mu}X^I\g^{\mu}\G^I\theta +  
\ternary{X^IX^JX^K}\G^{IJK}\theta\\
\label{delA}
\delta A_{\mu}(\Phi) = i\bar{\theta}\g_{\mu}\G^I\Omega_{\psi
X^I}(\Phi),
\end{gather} 
where $\Phi$ is either of the bosonic or fermionic fields, $X^I$ or
$\psi$. Let us note that \eq{susypsi} can be re-written in terms of
$\Omega$ as
\begin{equation} 
\label{psiomega}
\delta \psi = D_{\mu}X^I\g^{\mu}\G^I\theta +  
\frac{1}{3!}\Omega_{X^IX^J}(X^K)\G^{IJK}\theta.
\end{equation} 

The expression on the right hand side of \eq{delA} requires
qualification in the case when $\Phi$ is the fermionic field.
In a product of two fermions $\bar\psi$ must
appear on the left of $\psi$, in order to combine into a scalar in the
fermionic indices. Thus, when $\Phi=\psi$, we need to arrange the
fermions on the right hand side of \eq{delA} such that the fermions
are in proper order. 
Accordingly, in this case, using the following expressions
\begin{equation}
\label{fermind}
\begin{split}
\ternary{(\bar{\theta} \g_{\mu} \G^I \psi)~ X~ (\bar{\theta} \G^J
\psi) }
= -\ternary{(\bar{\psi} \g_{\mu} \G^I \theta)~ X~ (\bar{\theta} \G^J \psi)}\\
\ternary{(\bar{\theta} \G^I \psi)~ X~ (\bar{\theta}\g_{\mu} \G^J
\psi)} = 
- \ternary{(\bar{\psi}\G^I \theta)~ X~ (\bar{\theta}\g_{\mu} \G^J
\psi)}
\end{split}
\end{equation}
whenever necessary, the expression  on the right hand side of
\eq{delA} becomes 
\begin{equation}
\delta A_{\mu}(\psi) = -iM_{\nu}(\psi)
\g^{\nu}\g_{\mu}\G^I\theta,
\end{equation} 
where 
\begin{equation} 
\label{Mnu}
M_{\nu}(\Phi)=
\langle \bar{\psi}\g_{\nu}\dual{\psi}\Phi \rangle 
- \langle \bar{\psi}\dual{\Phi}\g_{\nu}\psi \rangle+ \langle \Phi\dual{\bar{\psi}}\g_{\nu}\psi \rangle.
\end{equation} 
The  Fierz identities,
\begin{gather}
\label{fierz}
(\bar{\psi} \chi)\lambda = -\frac{1}{2}(\bar{\psi}\lambda)\chi - \frac{1}{2}(\bar{\psi}\g_{\mu}\lambda)\g^{\mu}\chi \\
\lambda(\bar{\psi} \chi) = -\frac{1}{2}\psi(\bar{\lambda} \chi) +\frac{1}{2}\g_{\mu}\psi(\bar{\lambda}\g^{\mu} \chi)
\end{gather}
where $\psi$, $\chi$ and $\lambda$ are sixteen-dimensional Majorana
spinors, were used to prove the supersymmetry-invariance of the action
\eq{terac}.

Commutators of the supersymmetry transformations furnish
consistent equations of motion, as well as the gauge transformation.
For the bosonic fields, $X$, the commutator of two supersymmetry
transformations $\delta_1$ and $\delta_2$ with supersymmetry
parameters $\theta_1$ and $\theta_2$, respectively, take the form
\begin{equation}
\label{commX}
[\delta_1,\delta_2] X^I  = v^{\mu}D_{\mu}X^I + \Lambda(X^I)
\end{equation} 
where 
\begin{gather} 
v^{\mu} = -2i\bar{\theta}_2\g^{\mu}\theta_1\\
\Lambda = i\bar{\theta}_2\G^{JK}\theta_1\Omega_{X^JX^K}.
\end{gather} 
The second term in \eq{commX} is to be interpreted as a gauge
transformation. Considering the most general gauge transformation of
this kind, let us consider 
\begin{equation}
\Lambda_{fg}(X) = \Omega_{fg}(X),
\end{equation} 
for arbitrary functions $f,g$ on the world-volume. 
Let us note that $\Omega$ is anti-symmetric in the lower two indices
by definition, \eq{omegadef}. Using the definition of the vector
field \eq{Amu} and the derivative \eq{deriv}, then, the latter is
covariant, that is $\Lambda_{fg}(DX)=D(\Lambda_{fg}(X))$, 
if $A_{\mu}^{ab}$ transforms as
\begin{equation}
A_{\mu}^{\prime}(\Omega_{fg}(X)) = 
\frac{1}{2}(\Omega_{\pa_{\mu}f,g}(X)+
\Omega_{f,\pa_{\mu}g}(X)) +\Omega_{fg}(A_{\mu }(X)).
\end{equation} 
The commutator acting on  the fermion yields
\begin{equation} 
\label{commpsi}
\begin{split}
[\delta_1,\delta_2]\psi = v^{\mu} D_{\mu}\psi &+ \Lambda(\psi)\\
&+i \bar{\theta}_2\g^{\nu}\theta_1\g_{\nu}
\left(\g^{\mu}
D_{\mu}\psi-\frac{1}{2}\G^{IJ}\Omega_{X^IX^J}(\psi)\right)\\
&-\frac{i}{4}\bar{\theta}_2\G^{MN}\theta_1 \G^{MN} 
\left(\g^{\mu}
D_{\mu}\psi-\frac{1}{2}\G^{IJ}\Omega_{X^IX^J}(\psi)\right).
\end{split}
\end{equation} 
Closure of the supersymmetry algebra 
requires the braced expression in the second and the third terms
on the right hand side to vanish,
\begin{equation}
\label{eqpsi}
\g^{\mu}
D_{\mu}\psi-\frac{1}{2}\G^{IJ}\Omega_{X^I X^J}(\psi) = 0.
\end{equation}
This being the equation of motion obtained by varying $\psi$ from
\eq{terac}, we conclude that the algebra closes on-shell. 

Closure of the supersymmetry algebra on the vector field imposes
further restriction on the ternary bracket. 
The gauge field acted on by the same commutator leads to  
terms proportional to the rank five Gamma matrix, $\G^{IJKLM}$. The
coefficient of proportionality is to vanish for the closure of the
algebra. We find 
\begin{equation}
[\delta_1,\delta_2]A_{\mu}(\Phi) =
2i\epsilon_{\mu\nu\lambda}(\bar{\theta}_2\g^{\lambda}\theta_1)
\left(\Omega_{D_{\nu}X^I,X^I}(\Phi)-i M_{\nu}(\Phi) \right)
+D_{\mu}\Lambda(\Phi),
\end{equation} 
where $\Phi$ is either a bosonic field, $X^I$, or the fermion,
$\psi$  and 
$M_{\nu}(\Phi)$ is defined in \eq{Mnu}, provided we assure the
vanishing of the coefficient of $\G^{IJKLM}$ by imposing the 
condition
\begin{equation}
\label{constrt}
\begin{split}
&\left[\ternary{\Phi X^I\ternary{X^J X^K X^L}} 
+\ternary{X^I\ternary{X^J X^K X^L}\Phi} 
+\ternary{\ternary{X^J X^K X^L}\Phi X^I}\right.\\
&\left.\;-\ternary{X^I\Phi \ternary{X^J X^K X^L}} 
-\ternary{\Phi\ternary{X^J X^K X^L}X^I} 
-\ternary{\ternary{ X^J X^K X^L}X^I\Phi}\right]_{IJKL} =0,
\end{split}
\end{equation}
where $[\;]_{IJKL}$ designates complete
antisymmetrization  in
$\{I,J,K,L\}$. The constraint equation can be recast in terms of the
basis vectors of $V$ as
\begin{equation}
\label{algcstr}
\begin{split}
&\left[
\ternary{\tau_a\tau_b\ternary{\tau_c\tau_d\tau_e}} 
+\ternary{\tau_b\ternary{\tau_c\tau_d\tau_e}\tau_a} 
+\ternary{\ternary{\tau_c\tau_d\tau_e}\tau_a\tau_b } \right.\\
&\;\left. -\ternary{\tau_b\tau_a\ternary{\tau_c\tau_d\tau_e}} 
-\ternary{\tau_a\ternary{\tau_c\tau_d\tau_e}\tau_b} 
-\ternary{\ternary{\tau_c\tau_d\tau_e}\tau_b\tau_a } \right]_{bcde}=0.
\end{split}
\end{equation} 
Thus, the GJTS-II yields a consistent BLG theory if and only if the
algebra satisfies the constraint \eq{algcstr} in addition to \eq{JK1}
and \eq{JK2}. 
Equation \eq{constrt} is the most general constraint that can be written 
down, irrespective of the type of ternary algebra used. One can 
simplify it to the case of GJTS-II, by imposing the Jacobson and 
Kantor identities on \eq{constrt}.
Then the constraint takes the form 
\begin{equation}
\label{JKcont}
\begin{split}
&\left[\ternary{\Phi \ternary{X^J X^K X^L} X^I} - 
\ternary{\Phi X^I \ternary{X^J X^K X^L}}\right.\\
&-\left. \ternary{X^I X^K \ternary{X^J \Phi X^L}} - 
\ternary{X^J \ternary{X^K X^I \Phi} X^L}\right]_{IJKL} = 0.
\end{split}
\end{equation}

Let us point out that  instead of using the GJTS-II
if we assume the structure constants of
$\mathcal{A}$ to be  completely antisymmetric and impose the
fundamental identity, as in the traditional BLG theory, 
then the constraint \eq{algcstr} is satisfied. The present 
considerations generalize the BLG theory in this sense. 

Continuing with the genral structure, 
using $F_{\mu\nu} = [D_{\mu}, D_{\nu}]$ the supersymmetry
algebra closes up to the equation of motion for $A_{\mu}$, \viz
\begin{equation}
\label{fmunu}
F_{\mu\nu}   
-\epsilon_{\mu\nu\lambda}\left(\Omega_{D_{\lambda}X^I,X^I}(\Phi)+
i M_{\lambda}(\Phi)\right)  =0.
\end{equation} 

As a further check on consistency, let us point out that,
taking the supervariation of the equation of motion \eq{eqpsi} we get,
\begin{equation}
\label{bosgauge}
\begin{split}
&\G^I\left(D^2 X^I + iN_{IJ}(X^J) -\frac{1}{2} \Omega_{{X^J}{X^K}}
(\Omega_{{X^J}{X^K}}(X^I))\right)\theta \\
&- \g^{\lambda} \G^I \left(\frac{1}{2}\epsilon^{\mu \nu \lambda} 
F_{\mu \nu}(X^I) -\Omega_{D_{\lambda}X^J X^J}(X^I) - 
i M_{\lambda}(X^I)\right)\theta = 0.
\end{split}
\end{equation}
Here the second expression in braces is zero by \eq{fmunu}.
The remaining part gives the
equation of motion for the bosons,
\begin{equation}
\label{eqmotx}
D^2 X^I + iN_{IJ}(X^J) -\frac{1}{2}
\Omega_{{X^J}{X^K}}(\Omega_{{X^J}{X^K}}(X^I)) = 0,
\end{equation}
where we defined
\begin{equation}
\label{NX}
N_{IJ}(X^J) =\ternary{\bar{\psi}\G^{IJ}\psi X^J}
+\ternary{X^J \bar{\psi}\G^{IJ}\psi}
-\ternary{\bar{\psi} {X^J} \G^{IJ}\psi}.
\end{equation}
In here, too, the  quintic term coupled to the rank five gamma matrix cancels 
thanks to \eq{JK1}, \eq{JK2} and \eq{algcstr}.

We thus  conclude that the BLG theory can be consistently written 
in terms of a GJTS-II, provided the algebra satisfies  the constraint 
\eq{constrt}. 
\section*{Conclusion}
To summarize, in this note we consider the BLG theory of multiple 
M2-branes based on a
generalized Jordan triple system of the second kind. 
Unlike the prototypical
formulation of the BLG theory, the ternary bracket we use 
is not stipulated to  satisfy the so-called 
fundamental identity or complete antisymmetry of its structure
constants and thus evades the stringent restriction on
structure constants ensuing from it. 
Rather, the ternary product satisfies the 
Jacobson and Kantor identities, \eq{JK1} and \eq{JK2}, respectively, along with
an extra constraint, \eq{algcstr}. The traditional restrictions of
complete antisymmetry and fundamental identity turns out to be a
special solution to the constraint equation. In this sense, 
our considerations generalizes the BLG proposal. 
Therefore, the present formulation
is valid for a wide class of ternary algebras. 
It would be interesting to find out explicit ternary
algebras that is compatible with this formulation.

So far we have interpreted the constraint
\eq{algcstr} in the spirit of BLG 
as a condition on the ternary algebra itself assuming
the target to be flat, which severely
restrains the algebra. We can, instead, turn it around and interpret 
\eq{constrt} as constraining the $X$'s themselves, without restricting
the algebra. In this sense our formulation would give rise to
a system of M2-branes on a non-trivial target space.
In particular, 
considering representations of ternary algebras,
it would be illuminating to examine whether
a generic algebra can be incorporated in the scheme by
modifying the target space appropriately. 

One of the motivations for our  approach here is based on the fact that
a theory of multiple M2-branes, described by a ternary algebra is 
supposed to be
a strong-coupling limit of the theory of multiple D2-branes, the latter being  
described, in the low-energy regime,  
by a Yang-Mills theory.
This hints at a natural connection between a ternary
algebra and a Lie algebra. Since a Jordan triple system
of the second kind intrinsically leads to a graded Lie algebra,
namely, the Kantor algebra, the former is a likely candidate
to feature in a theory of M2-branes. 
However, the physical interpretation of the means of obtaining the
Kantor algebra from the GJTS-II within this context
is yet to be fleshed out. We hope to report on some of these issues in
future.   
\section*{Acknowledgment}
It is a pleasure to thank 
Shamik Banerjee,
Anirban Basu, 
Utpal Chattopadhyay, 
Bobby Ezuthachan, 
Pushan Majumdar
and Soumitra SenGupta
for very fruitful discussions at various points.
We are grateful to Jakob Palmkvist for pointing out an error in the 
normalization of the last term in \eq{terac} in the previous version. 

\end{document}